\documentclass[11pt]{article} 

\usepackage[utf8]{inputenc} 

\usepackage{geometry} 
\usepackage{graphicx} 

\usepackage{booktabs} 
\usepackage{array} 
\usepackage{paralist} 
\usepackage{verbatim} 
\usepackage{subfig} 

\usepackage{fancyhdr} 
\usepackage{sectsty}
\allsectionsfont{\sffamily\mdseries\upshape} 

\usepackage[nottoc,notlof,notlot]{tocbibind} 
\usepackage[titles,subfigure]{tocloft} 

\usepackage{indentfirst}

\title{\textbf{The utility of a convolutional neural network for generating a myelin volume index map from rapid simultaneous relaxometry imaging}}
\author{}
\date{} 
\usepackage{amsmath}

\begin{document}
\LARGE
\begin{center}
\textbf{The utility of a convolutional neural network for \\ generating a myelin volume index map from \\ rapid simultaneous relaxometry imaging}
\end{center}

\vspace{0.5cm}

\large
{\setlength\leftskip{1.0cm}
\noindent
Yasuhiko Tachibana$^{1,2}$, Akifumi Hagiwara$^{2,3}$, Masaaki Hori$^{2}$, Jeff Kershaw$^{1}$, Misaki Nakazawa$^{2}$, Tokuhiko Omatsu$^{1}$, Riwa Kishimoto$^{1}$, Kazumasa Yokoyama$^{4}$, Nobutaka Hattori$^{4}$, Shigeki Aoki$^{2}$, Tatsuya Higashi$^{1}$, and Takayuki Obata$^{1}$

}

\small
\vspace{0.5cm}
{\setlength\leftskip{1.0cm}
\noindent
1. Department of Molecular imaging and Theranostics, National Institute of Radiological Sciences, QST, 4-9-1, Anagawa, Inage-ku, Chiba, Japan, 263-8555

\noindent
2. Department of Radiology, Juntendo University School of Medicine, 2-1-1, Hongo, Bunkyo-ku, Tokyo, Japan, 113-8421

\noindent
3. Department of Radiology, Graduate School of Medicine, The University of Tokyo, 7-3-1, Hongo, Bunkyo-ku, Tokyo, Japan, 113-8655

\noindent
4. Department of Neurology, Juntendo University School of Medicine, 2-1-1, Hongo, Bunkyo-ku, Tokyo, Japan, 113-8421

\vspace{0.5cm}
\noindent
Email: tachibana.yasuhiko@qst.go.jp

}

\subsection*{}

{\setlength\leftskip{2.0cm}
\noindent
\textbf{Abstract}

\noindent
\textbf{Background and Purpose:}
A current algorithm to obtain a synthetic myelin volume fraction map (SyMVF) from rapid simultaneous relaxometry imaging (RSRI) has a potential problem, that it does not incorporate information from surrounding pixels. The purpose of this study was to develop a method that utilizes a convolutional neural network (CNN) to overcome this problem.

\noindent
\textbf{Methods:}
RSRI and magnetization transfer images from 20 healthy volunteers were included. A CNN was trained to reconstruct RSRI-related metric maps into a myelin volume-related index (generated myelin volume index: GenMVI) map using the myelin volume index map calculated from magnetization transfer images (MTMVI) as reference. The SyMVF and GenMVI maps were statistically compared by testing how well they correlated with the MTMVI map. The correlations were evaluated based on: (i) averaged values obtained from 164 atlas-based ROIs, and (ii) pixel-based comparison for ROIs defined in four different tissue types (cortical and subcortical gray matter, white matter, and whole brain). 

\noindent
\textbf{Results:}
For atlas-based ROIs, the overall correlation with the MTMVI map was higher for the GenMVI map than for the SyMVF map. In the pixel-based comparison, correlation with the MTMVI map was stronger for the GenMVI map than for the SyMVF map, and the difference in the distribution for the volunteers was significant (Wilcoxon sign-rank test, P$<$.001) in all tissue types. 

\noindent
\textbf{Conclusion:}
The proposed method is useful, as it can incorporate more specific information about local tissue properties than the existing method.

}

\normalsize
\section{Introduction}
Measuring myelin volume using MRI is useful for evaluating the development and aging of humans and for assessing the progression of degenerative or demyelinating diseases [1-3]. However, MRI-based methods remain problematic because none of the existing techniques can replace pathological measurement, and moreover, its lengthy scan-time [1].

Recently, rapid simultaneous relaxometry imaging (RSRI) was developed to measure the longitudinal relaxation rate (R1), transverse relaxation rate (R2), proton density (PD), and the local B1 field from a single scan within an acceptable scan-time [4]. The metrics can be used to estimate the myelin volume fraction using a commercial software package (SyMRI) [5]. The myelin volume fraction estimated with SyMRI (i.e. SyMVF) was reported as being highly correlated with a widely used measure of myelin obtained from magnetization-saturation imaging (MTsat) [6, 7], namely the magnetization transfer-based myelin volume index (MTMVI) [1]. The usefulness of SyMVF has already been noted in several applications [8-11]. However, there might be room for improvement in the algorithm used to estimate a SyMVF map. The SyMVF metric is determined pixel-by-pixel from a lookup-table that connects combinations of R1, R2, and PD values to a myelin volume fraction [9], meaning that a pixel with a certain combination of R1, R2, and PD values is always assigned the same myelin volume fraction without considering any local properties. As tissue structure differs in different areas of the brain (e.g. neuron count [12], neuron fiber radius [12, 13], iron deposition), this could lead to inaccuracy when generating the SyMVF map. Adding information about local tissue properties may strengthen the accuracy and reliability of the output myelin volume fraction map.

Recently, the convolutional neural network (CNN) technique achieved great success for image segmentation of many areas in the human body [14-17]. As the shape of the data of each layer of a CNN is generally unrestricted, taking the processing stream from one CNN and adding it to the function of another CNN is possible (e.g. [15-17]). Based on this idea, this study combines a CNN for segmentation with another simple CNN designed for general non-linear reconstruction so that the finally generated myelin volume index (GenMVI) is more specific to the characteristics of the tissue in each pixel. 

The purpose of this study was to evaluate the usefulness of this method for estimating myelin-volume in human brain.

\section{Materials and Methods}

 This is a retrospective study. The data used in this study was originally acquired for another previous study [1]. This study was approved by the IRB of Juntendo University. Written informed consent was obtained from all participants.

\subsection{Study participants}

Twenty healthy volunteers, nine males (25 to 67 years, mean 53.2 years) and eleven females (44 to 71 years, mean 57.0 years), without neurological or psychological history, were included as subjects. Images acquired from the subjects were screened by two board-certified radiologists (Y.T. and A.H., 12 and 5 years of experience interpreting brain MRI, respectively) to confirm that no moderate-to-severe white-matter ischemic lesions (Fazekas grade 2 or more [18]), asymptomatic cerebral infarction, or regional brain atrophy existed.

\subsection{Image acquisition and data processing to generate SyMVF and MTMVI maps}

All scans in this study were performed by 3T MRI scanner (MAGNETOM Prisma, Siemens Healthcare, Erlangen) using a 64-channnel head coil. Images were acquired using the QRAPMASTER imaging sequence [1], which has two different TEs (i.e. 22 and 99 ms) and four different saturation delay times (i.e. 170, 620, 1970, and 4220 ms) in a single scan. The other major parameters for QRAPMASTER were: TR 4250 ms; field of view 230$\times$186 mm; matrix 320$\times$186; slice thickness /gap 4.0 /1.0 mm. The acquired images were processed using SyMRI 8.0 software (SyntheticMR, Linkoping, Sweden) to obtain R1, R2, PD, and SyMVF maps. A brain-area probability map (BAP) and synthetic T1-weighted image were also automatically generated during this process.

Imaging with magnetization-transfer (MT) weighting was performed as a first step to obtain MTMVI images. Images were acquired using FLASH sequence with T1-, PD-, and magnetization transefer-weightings. TR and excitation flip angle were set at 10 ms and 13 degrees for T1-weighted images, and 24 ms and 4 degrees for PD- and MT-weighted images. An off-resonance Gaussian-shaped RF pulse (frequency offset from water resonance 1.2 kHz, pulse duration 9.984 ms, and nominal flip angle 50 degrees) was adopted for the MT-weighted images. The other major parameters were: field-of-view 224$\times$224 mm; matrix 128$\times$128; slice thickness 1.8 mm. The MTsat map was calculated from the images as described in a previous report [1], and it was then scaled as described in the next two sections to generate a final MTMVI map.

\subsection{ROI definition}

The Johns Hopkins University (JHU) ICBM-DTI-81 WM labels atlas [19, 20] and Automated Anatomical Labeling (AAL) atlas [21, 22] were used to define 48 local ROIs for the WM area, and 108 and 8 ROIs for the cortical and subcortical GM areas, respectively.

Local ROIs of the atlases were registered to SyMVF volumes for each volunteer. First, the synthetic T1-weighted image volume of a volunteer was registered to the MNI152 template using the FMRIB Software Library (FSL) linear and non-linear image registration tools (FLIRT and FNIRT) [23, 24]. The warp function defined in this registration was then inverted to warp the atlas ROIs (total 164 local ROIs) to fit the volunteer’s SyMVF space. In addition, the warped ROIs were grouped and merged to form another set of ROIs: cortical GM (ROI$_{\text{cGM}}$), subcortical GM (ROI$_{\text{sGM}}$) and WM (ROI$_{\text{WM}}$). These three ROIs were eroded once with an eight-connected-neighborhood rule to avoid partial-volume effects at the margins of each tissue type. Note that when using one of the 164 local ROIs in the analysis, pixels within the ROI that had been eroded from either ROI$_{\text{cGM}}$, ROI$_{\text{sGM}}$, or ROI$_{\text{WM}}$ were removed for the same reason. In addition, a whole brain ROI (ROI$_{\text{WB}}$) was created by merging ROI$_{\text{cGM}}$, ROI$_{\text{sGM}}$, and ROI$_{\text{WM}}$. Furthermore, local ROIs corresponding to the genu, body, and splenium of the corpus-callosum were merged to form a single ROI for the corpus-callosum (ROI$_{\text{CC}}$).

ROI definition for the MTsat volume was performed in the same way as for the SyMVF map, except that a 3D T1-weighted image volume was used instead of a synthetic T1-weighted image volume.

\subsection{MTMVI maps}

MTsat volumes were scaled to create the MTMVI map so that all myelin-related map images in this study are on the same scale. For each MTsat and SyMVF volume pair, pixels in the WM area were extracted and averaged using the ROI$_{\text{WM}}$ defined for each volunteer in the previous section. The MTMVI volume was created by multiplying MTsat volume by a constant scaling value so that the average values of pixels in ROI$_{\text{WM}}$ were equal for the SyMVF and MTMVI volumes. 

Finally, each MTMVI volume was nonlinearly registered to the SyMVF volume space of the same volunteer. The Advanced Normalization Tools (ANTS, antsRegistrationSyNQuick.sh, http://stnava.github.io/ANTs/) package [25] was used for this purpose.

\subsection{Deep-learning-based method to obtain myelin volume index}

\subsubsection{Image preparation}

Computing procedures described in this section were performed using our in-house software running on MATLAB 2017b® (Mathworks, Natick). To create a dataset for training, first, 32$\times$32-pixel patch images were randomly subsampled from each slice of all image volumes (i.e. R1, R2, PD, BAP, SyMVF, and MTMVI) corresponding to each volunteer. Patches including brain area (defined from BAP map as pixels with probability $>$ 0.95) of less than half of the whole area were excluded. Finally, approximately 6000 patch sets were subsampled from each volunteer. All subsampled patches were resized to 128$\times$128.

To create another dataset for testing, a similar subsampling procedure was repeated for each volunteer. For testing data, the patches were not randomly subsampled but regularly in five-pixel strides.

\subsubsection{CNN training and generating GenMVI volumes}

The designed CNN network consists of a segmentation block for acquiring local information from R1, R2, and PD maps, and a reconstruction block for adding that information to the corresponding SyMVF map. Details are described in Figure 1.

\begin{figure}
	\centering
	\includegraphics[width=15cm]{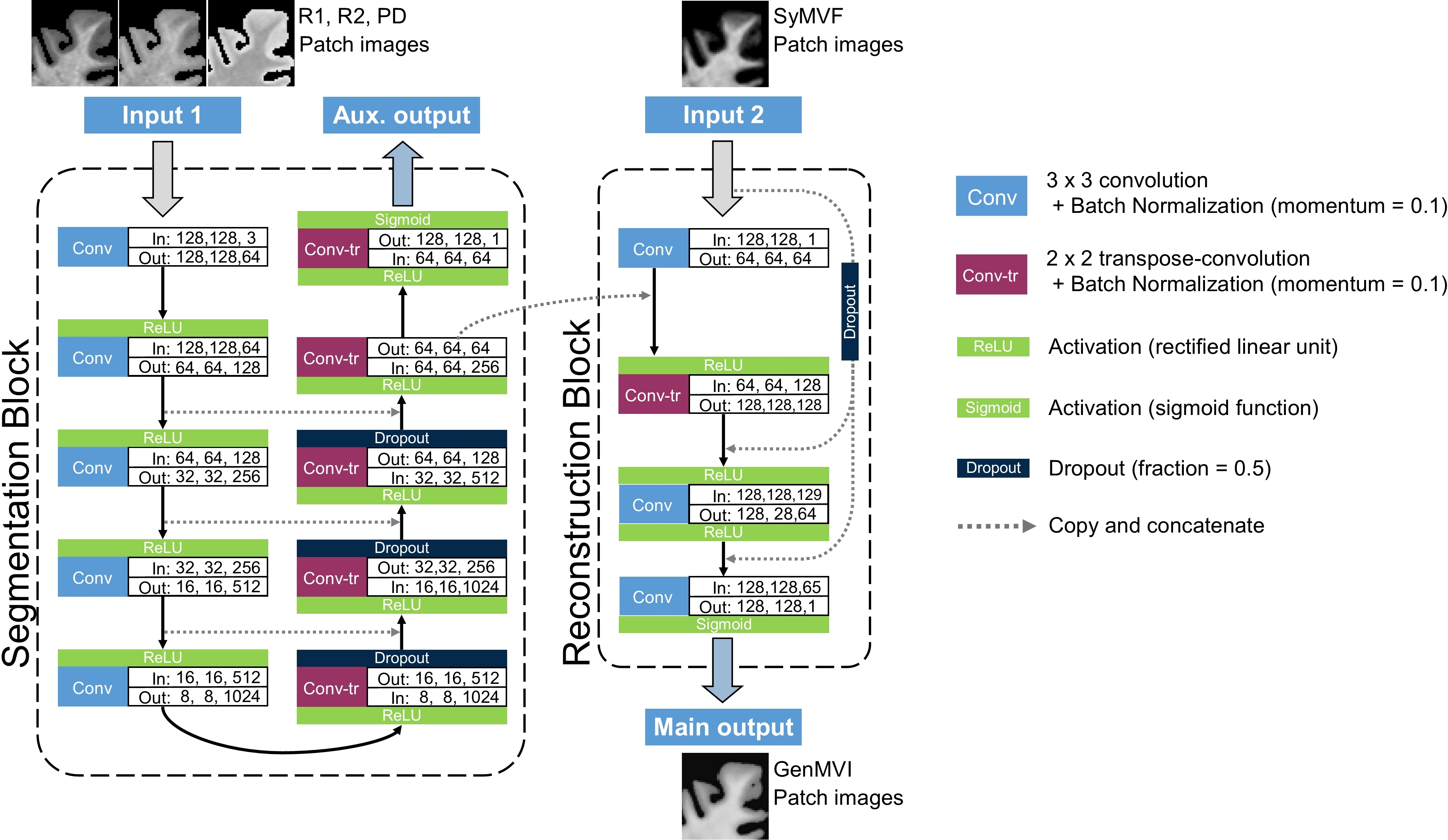}
	\caption {\textbf{The network architecture designed for this study.} The network consisted of a segmentation block and a reconstruction block. The segmentation block has a contracting pathway on the left side and an expanding pathway on the right side. The block was designed to extract local information from the R1, R2, and PD maps. The reconstruction block reconstructs the corresponding SyMVF map into a new map image (generated myelin volume index: GenMVI). The numbers indicated at each convolutional (Conv) or transpose-convolutional (Conv-tr) layer describe the size of the input and output images (rows, columns, and channels) for the layer.}
	\label{Figure1}

\end{figure}

To create a GenMVI map for a particular volunteer, data from the other 19 were used to train a CNN for that volunteer (leave-one-out cross-validation). The 19 volunteers were randomly assigned to two groups of 15 and 4 subjects to create training and validation data sets. Gaussian noise was added to the training data to avoid overfitting.

Training was performed using the Tensorflow-gpu (version 1.8.0) [26] platform with Keras [27] (version 2.1.6). Calculations were performed by computer equipped with: dual CPU, Intel Xeon® E5-2623v4; dual GPU, Nvidia TitanX® Pascal, 12GB GDDR5X; 128 GB random access memory; Ubuntu 16.04LTS. R1, R2, and PD patches were input to the segmentation block, and the SyMVF patches were input to the reconstruction block. As loss-function, the root-mean-square-error (RMSE) with respect to MTMVI was obtained for both main output and auxiliary outputs, and then summed after multiplying RMSE of the auxiliary output by 0.2. The Adam algorithm [28] was applied for optimization, where the learning rate started from 0.0001 at the first epoch and then decreased according to the hyperbolic function,

\[lr(n) = \frac{tanh(1.8-0.3n)+1}{2(tanh(1.5)+1)}  \]

where $lr(n)$ is the learning rate for epoch number $n$. The maximum number of epochs for training was set at 10, but training was aborted when the loss obtained at the end of each epoch using the validation data did not decrease for three consecutive epochs.

After completion of training, a GenMVI map was generated for each volunteer by test dataset. The dataset for each volunteer was input to a trained CNN (i.e. trained by the other 19 volunteers for each volunteer) to obtain main output as GenMVI patches. The output patches were re-orientated to form a whole GenMVI volume.

\subsection{Statistical analysis}

\subsubsection{Analysis based on averaged value of pixels inside ROIs}

The 164 local ROIs were separately applied to the MTMVI, SyMVF, and GenMVI maps of the volunteers. The values of the pixels included in each ROI were averaged and recorded. The absolute difference of the averaged values of the SyMVF and MTMVI maps were calculated for each local ROI ($\Delta$Sy), and the same was done for the GenMVI and MTMVI maps ($\Delta$Gen). $\Delta$Sy and $\Delta$Gen were statistically compared for each of the following four regions: (i) cortical GM (consisting of 108 ROIs), (ii) subcortical GM (consisting of 8 ROIs), (iii) WM (consisting of 48 ROIs), and (iv) whole brain (consisting of all 164 ROIs). The Wilcoxon signed-rank test was applied and P$<$.05 was considered significant.

For further comparison, Pearson’s correlation analysis was performed between MTMVI and SyMVF, and between MTMVI and GenMVI for the averaged values calculated for a total of 3280 local ROIs (164 ROIs from each of the 20 volunteers).

\subsubsection{Pixel-based comparison within ROIs}

ROI$_{\text{cGM}}$, ROI$_{\text{sGM}}$, ROI$_{\text{WM}}$, and ROI$_{\text{WB}}$ were applied to each volunteer. The pixels in these four ROIs were extracted and used to calculate a pixel-based Pearson’s correlation coefficient for both SyMVF and GenMVI in comparison to MTMVI, for each volunteer. Distributions of SyMVF-based and GenMVI-based correlation coefficients for the 20 volunteers were compared statistically for all four ROIs. Wilcoxon signed-rank test was applied for this purpose and P$<$.05 was considered significant.

In addition, a similar pixel-based comparison was performed for ROI$_{\text{CC}}$.

\section{Results}

Examples of MTMVI, SyMVF, and GenMVI maps from the same volunteer are shown in Figure 2.

\begin{figure}
	\centering
	\includegraphics[width=15cm]{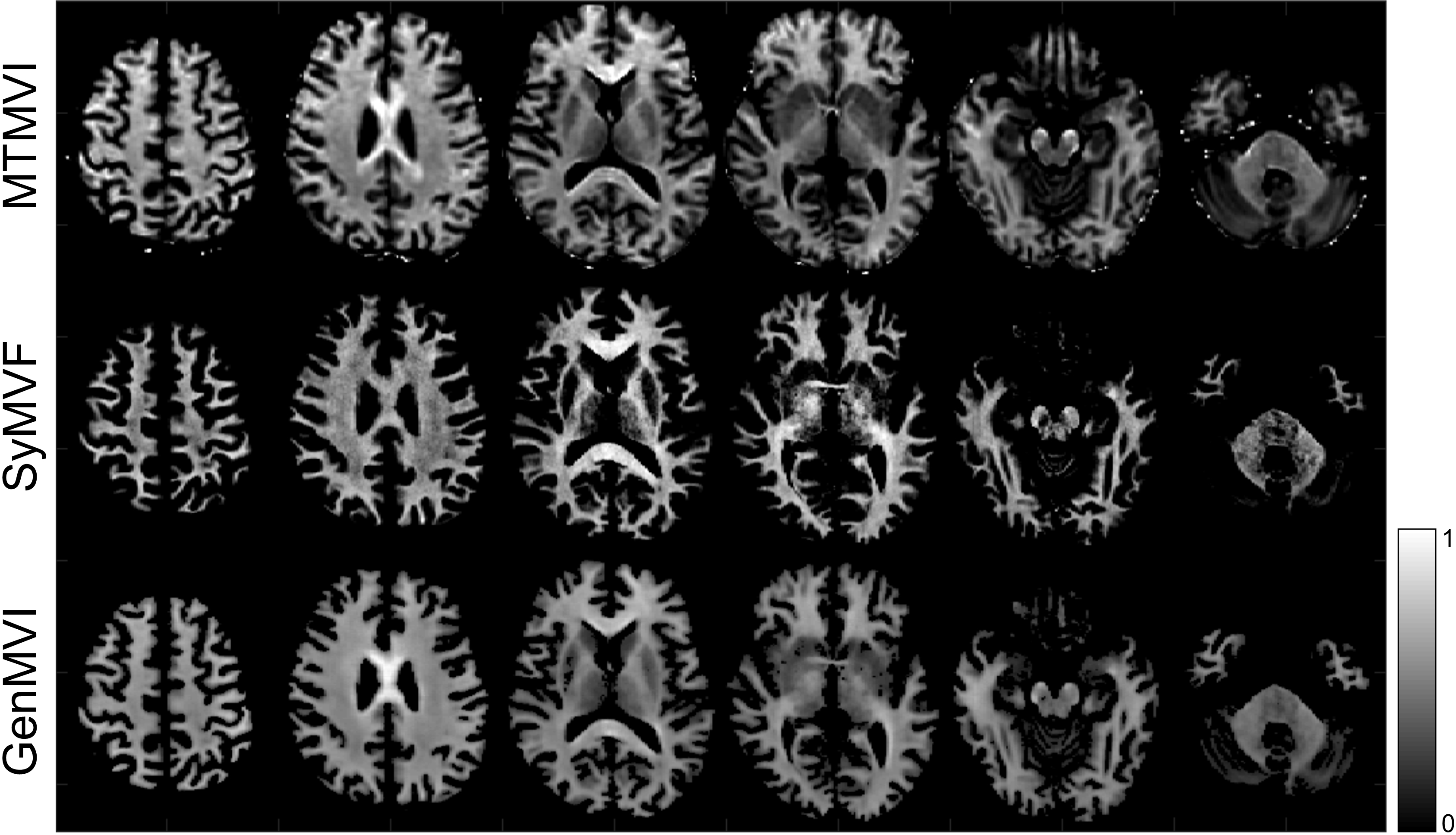}
	\caption {\textbf{Typical MTMVI, SyMVF, and GenMVI maps obtained from the same volunteer.} Visually, the contrast of the GenMVI map is more similar to MTMVI than that of the SyMVF map (See the corpus callosum area as a particular example). }
	\label{Figure2}

\end{figure}

\subsection{Results based on averaged value of pixels inside ROIs}

The distribution and relationship between the averaged values of the MTMVI, SyMVF, and GenMVI maps for the 164 local ROIs are illustrated in scatterplots (Figure 3). The median and minimum to maximum ranges of these metrics for cortical GM, subcortical GM, WM, and whole brain areas are indicated in Table 1. The median and range of the absolute differences with respect to MTMVI (i.e. $\Delta$Sy and $\Delta$Gen) are also shown in Table 1. The median value of $\Delta$Gen was smaller than that of $\Delta$Sy for all areas, with all differences being significant (P$<$.001). Pearson’s correlation coefficient obtained from all local ROIs from the 20 volunteers was larger for MTMVI and GenMVI (R=0.86) than for MTMVI and SyMVF (R=0.77) (Figure 3). Both correlations were significant (P$<$.001).

\begin{figure}
	\centering
	\includegraphics[width=15cm]{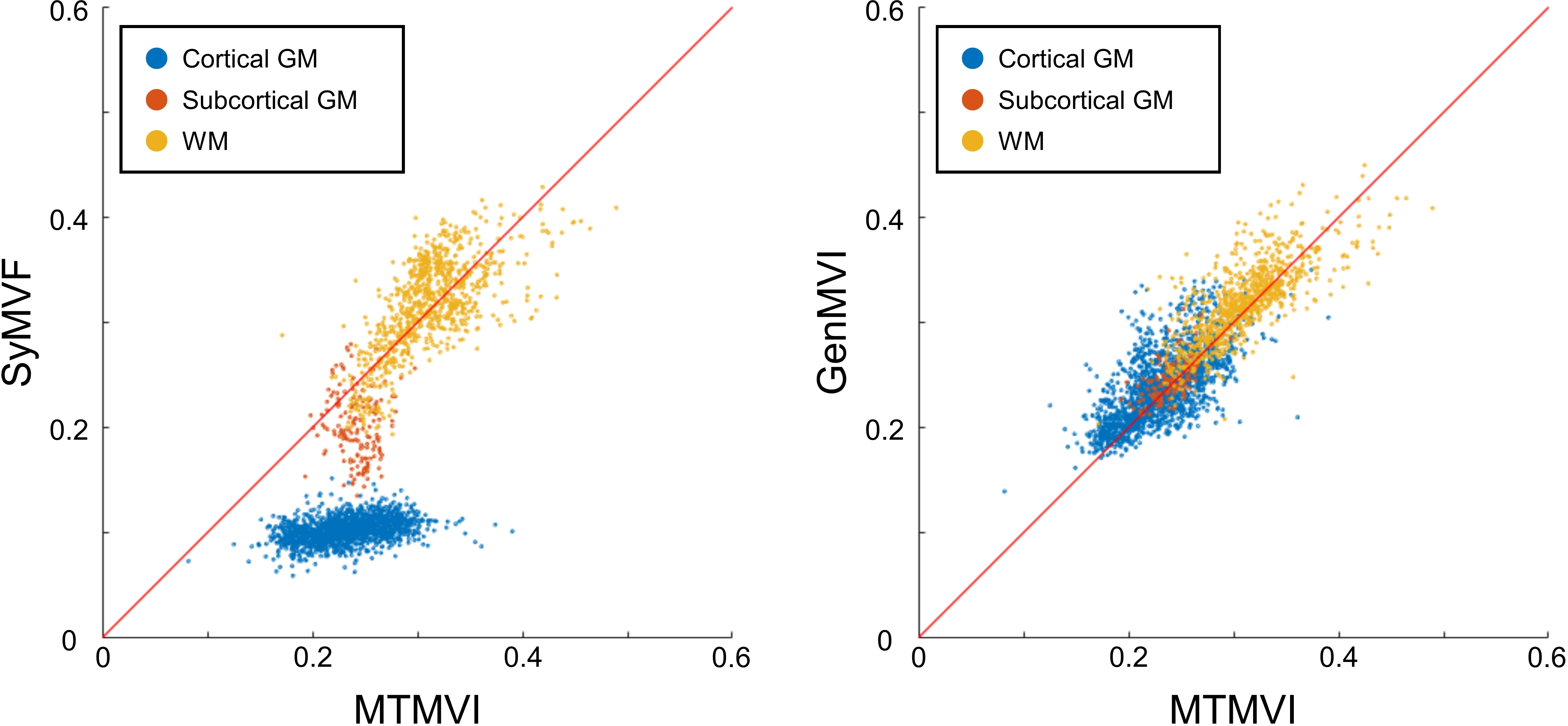}
	\caption {\textbf{Scatter plots of the average values in the 164 local ROIs:} correlation between (left) the MTMVI and SyMVF maps (R=0.77), and (right) the MTMVI and GenMVI maps (R=0.86). Overall, the values are more consistent between maps for the latter comparison. }
	\label{Figure3}

\end{figure}

\begin{table}[b]
\begin{minipage}{\textwidth}
\caption{\textbf{The absolute errors between the SyMVF and MTMVI maps ($\Delta$Sy), and between the GenMVI and MTMVI maps ($\Delta$Gen)}}
\label{Table1}
\begin{center}
	\begin{tabular}{lcccc}
					& cortical GM 	& Subcortical GM	& WM 			& Whole Brain 	\\ \hline
	MTMVI			& 0.23 [0.08, 0.39]	& 0.24 [0.19, 0.30]	& 0.32 [0.17, 0.49]	& 0.25 [0.08, 0.49]	\\
	SyMVF			& 0.10 [0.06, 0.15]	& 0.20 [0.14, 0.28]	& 0.32 [0.19, 0.43]	& 0.11 [0.06, 0.43]	\\
	GenMVI			& 0.24 [0.14, 0.35]	& 0.25 [0.21, 0.31]	& 0.32 [0.20, 0.45]	& 0.26 [0.14, 0.45]	\\
	$\Delta$Sy		& 0.13 [0.00, 0.29]	& 0.04 [0.00, 0.11]	& 0.02 [0.00, 0.12]	& 0.10 [0.00, 0.29]	\\
	$\Delta$Gen		& 0.02 [0.00, 0.15]	& 0.01 [0.00, 0.06]	& 0.01 [0.00, 0.11]	& 0.02 [0.00, 0.15]	\\
	P-value			& $<$.001		& $<$.001		& $<$.001		& $<$.001 \\ \hline
	\end{tabular}
\end{center}
\begin{flushright}
Results are presented as median [min, max].
\end{flushright}
\end{minipage}
\end{table}

\subsection{Results of pixel-based comparison within ROIs}

Distributions of pixel-based correlation-coefficients for the 20 volunteers for each of ROI$_{\text{cGM}}$, ROI$_{\text{sGM}}$, ROI$_{\text{WM}}$, and ROI$_{\text{WB}}$ are illustrated in box-plot graphs (Figure 4). Strong correlation between MTMVI and GenMVI was found for all four ROIs, where median values of the correlation coefficients were always higher than 0.80. On the other hand, those for SyMVF were lower than 0.7 except for ROI$_{\text{WM}}$, for which the value was 0.70. Distributions for SyMVF and GenMVI were significantly different for all four ROIs (P$<$.001).

The result of an additional pixel-wise comparison for ROI$_{\text{CC}}$ is also illustrated in a box-plot graph (Figure 5). The correlation was moderate for GenMVI (median value 0.56), but it was stronger than that for SyMVF (median value 0.21). The difference in distributions was significant (P$<$.001).

\begin{figure}
	\centering
	\includegraphics[width=15cm]{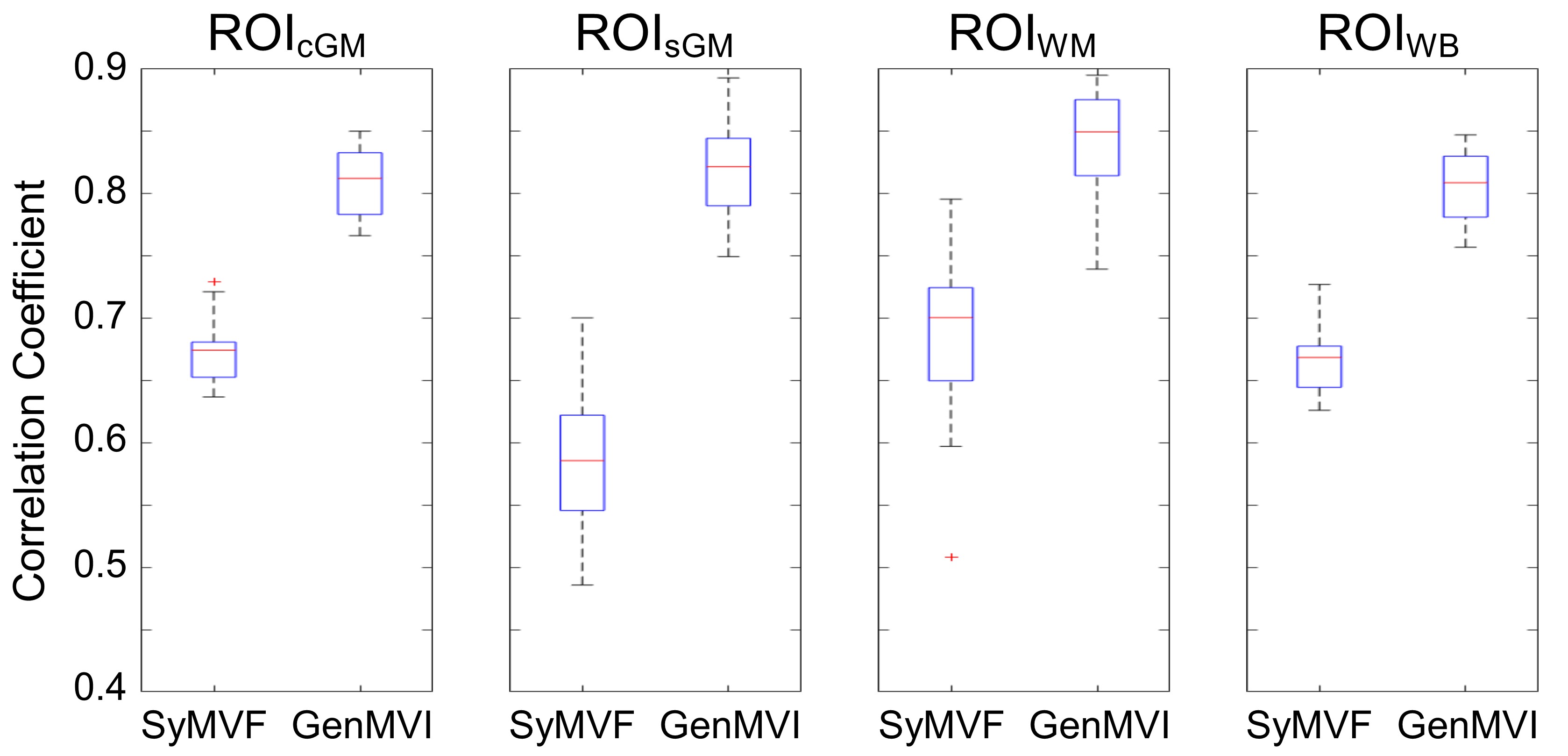}
	\caption {\textbf{The box plots show the distribution of pixel-wise correlation coefficients obtained by comparing the SyMVF and GenMVI maps with the MTMVI map for the four ROIs corresponding to the cortical GM, subcortical GM, WM, and whole brain (i.e. ROI$_{\text{cGM}}$, ROI$_{\text{sGM}}$, ROI$_{\text{WM}}$, and ROI$_{\text{WB}}$), for all 20 volunteers.} The median values are higher for the GenMVI map than for the SyMVF map, and the differences of the distributions are significant in all four areas (Wilcoxon signed-rank test, P$<$.001). }
	\label{Figure4}
		
\end{figure}

\begin{figure}
	\centering
	\includegraphics[width=4cm]{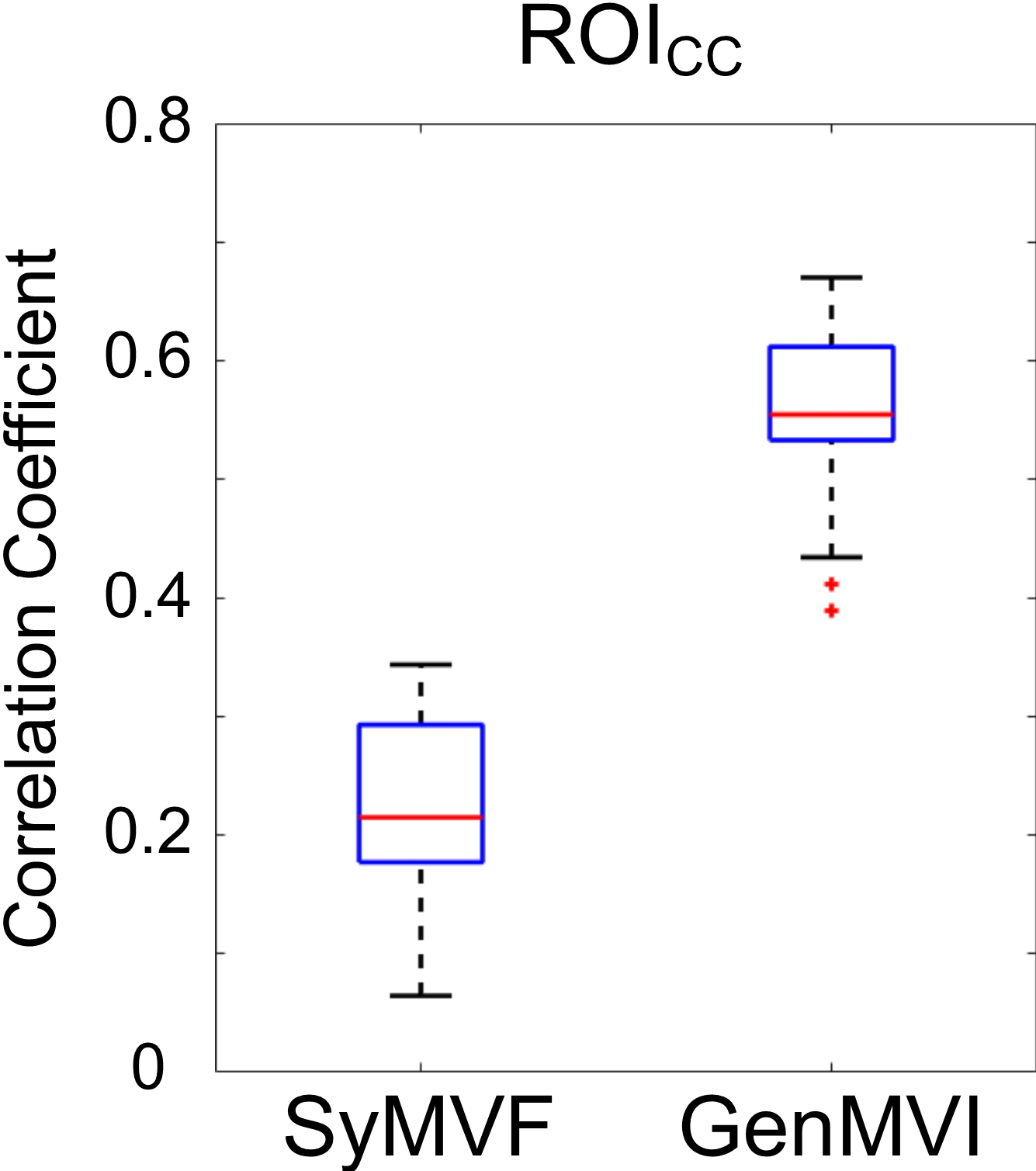}
	\caption {\textbf{The box plots show the distribution of pixel-wise correlation coefficients obtained by comparing the SyMVF and GenMVI maps with respect to the MTMVI map using the corpus-callosum ROI (ROI$_{\text{CC}}$).} The correlation coefficient obtained for a GenMVI map is always higher than that obtained for the corresponding SyMVF map. The difference is significant (P$<$.001, Wilcoxon signed-rank test).}
	\label{Figure5}

\end{figure}

\section{Discussion}

This study was designed to evaluate the usefulness of a CNN-based method for estimating a myelin-volume index map from RSRI images. Combining the overall results, it seems that the method proposed in this study successfully reconstructs the contrast of a SyMVF map of healthy brain into a new contrast that is more strongly related to the corresponding MTMVI map.

\subsection{CNN architecture}

The segmentation block aimed to compensate for the potential weak point of a SyMVF map, namely, not containing any information about local tissue structure. The segmentation block consists of many convolutional layers between input and output, which means that the value assigned to a pixel at the output incorporates information from a relatively wide area surrounding the pixel in the input image (i.e. maximum 32$\times$32-pixel area). This block was designed based on U-net [14], which is a network that has achieved great success when used to segment various anatomical and histological images [29-31].

The overall CNN designed in this research aimed to utilize the high capability of CNN for flexible image segmentation and reconstruction, while at the same time designing the network so that the priority of the SyMVF map as input is relatively high (e.g. shortcuts in reconstruction block that sent SyMVF images to later layers without being merged with the input from the segmentation block). The design aimed to lessen the black box problem [32, 33] as much as possible by using SyMVF, which has a logical foundation (i.e. Bloch simulation), as basic starting point and giving more priority in the overall network design.

\subsection{Comparison of SyMVF and GenMVI with respect to MTMVI}

Visually, the contrast of the GenMVI map was closer than the contrast of SyMVF to that of MTMVI (Figure 2). In particular, characteristics of the corpus callosum were better reproduced in the GenMVI map than in the SyMVF map. The corpus callosum was where the contrast was especially different between the SyMVF and MTMVI maps in a previous study [1]. The method proposed in this study appears to have improved this problem.

From analysis using the averaged values of the 164 local ROIs, values of the GenMVI map were more similar to the MTMVI map for different brain areas than those of the SyMVF map (Figure 3, Table 1). In addition, even though the correlation for SyMVF was also strong (R=0.77), which is consistent with a previous study [1], the overall linear correlation with MTMVI was higher for GenMVI (R=0.86) (Figure 3). 

The results of the pixel-based comparisons further support the results of the atlas-based comparison, namely, that the GenMVI map has a stronger correlation with MTMVI than the SyMVF map does (Figure 4). Pixel-based comparison was added because the results using averaged values for each ROI are potentially biased by a possible difference in pixel number in each ROI.

The pixel-based comparison performed for the corpus callosum indicated higher correlation for GenMVF over SyMVF, with a statistically significant difference between the two distributions (P$<$.001) (Figure 5). This result is consistent with the visual evaluation of the region (Figure 2).

Currently, one of the great advantages of SyMVF over simpler and less time-consuming myelin-related indices may be its high correlation with the widely used MTMVI map [1]. The proposed GenMVI maps may improve on the advantages of the SyMVF maps to expand the capability towards possible clinical use. However, MTMVI is still not a golden standard that is comparable to pathology. Thus, the fact that the contrast of the GenMVI map was closer than the SyMVF map to the MTMVI map does not directly mean that the GenMVI map is more accurate than the SyMVF map. Further study using pathology-based measures of myelin volume as a target during training is desirable. Moreover, since the GenMVI map has the limitation that its logical foundation is relatively weak, further clinical validation is also important.

A possible improvement for the proposed method would be to include new information in addition to the current input. The most promising candidates are various diffusion-related metrics [34, 35], such as apparent diffusion coefficient (ADC), fractional anisotropy (FA), and other metrics obtained from diffusional kurtosis imaging (DKI). It is well established that these parameters are closely related to the local micro-structure and myelin volume content in some tissues including demyelinating lesions [34]. The additional information may help to appropriately estimate myelin volume especially when the target will be expanded to pathological brains.

As a limitation of this study, the MTMVI maps were warped to register the images to R1, R2, PD and SyMVF maps obtained from RSRI. A small mis-registration might have affected the training, as well as the final results of the statistical analysis. 

\section{Conclusion}

In conclusion, the deep-learning-based method proposed in this study generated a myelin-volume index from RSRI that incorporates more specific information about local tissue properties than the existing technique. However, further work is necessary to validate the proposed method so that it might be employed for clinical use.

\section*{Acknowledgement}

The authors appreciate the assistance of Hiroko Kamada and Etsuko Mitsui during the study.

This work was supported by AMED under grant number JP18lk1010025; ImPACT Program of Council for Science, Technology, and Innovation (Cabinet Office, Government of Japan); JSPS KAKENHI grant number 17K10385; JSPS KAKENHI grant number 16K19852; JSPS KAKENHI grant number JP16H06280, Grant-in-Aid for Scientific Research on Innovative Areas– Resource and Technical Support Platforms for Promoting Research “Advanced Bioimaging Support”; and the Japanese Society for Magnetic Resonance in Medicine.

\end{document}